\documentstyle[12pt,epsf]{article}

\large
\newcommand{\be}{\begin{eqnarray}}
\newcommand{\ee}{\end{eqnarray}}

\begin{document}
\setlength{\baselineskip}{23pt}
\setlength{\baselineskip}{27pt}
\pagestyle{empty}
\renewcommand{\thefootnote}{\fnsymbol{footnote}}
\centerline{\bf\LARGE Phase Transitions at Finite Temperature and Dimensional}
\centerline{\bf\LARGE Reduction for Fermions and Bosons }
\vskip 1cm
\centerline{Aleksandar KOCI\' C and John KOGUT
}
\vskip 1cm
\centerline{\it Loomis Laboratory of Physics Physics}
\centerline{\it 1110 W. Green St.}
\centerline{\it University of Illinois}
\centerline{\it Urbana, Il 61801-3080}
\vskip 1cm

\setlength{\baselineskip}{16pt}
\centerline{\bf Abstract}

In a recent Letter we discussed the fact that
large-$N$ expansions and computer simulations indicate that the
universality class of the finite temperature chiral symmetry restoration
transition in the 3D Gross-Neveu model is mean field
theory. This was seen to be a counterexample to the standard 'sigma model'
scenario which predicts the 2D Ising model universality class. In this article
we present more evidence, both theoretical and numerical, that this result
is correct.
We develop a physical picture for our results and discuss the width of the
scaling region (Ginzburg criterion), $1/N$ corrections, and differences between
the dynamics of BCS superconductors and Gross-Neveu models.
Lattices as large as $12 \times 72^2$ are simulated for
both the $N=12$ and $N=4$ cases and the numerical evidence for mean field
scaling is quite compelling. We point out that the amplitude ratio for the
model's susceptibility is a particulartly good observable for
distinguishing between the dimensional reduction and the mean field scenerios,
because this universal quantity differs by almost a factor of $20$ in the two
cases. The simulations are done close to the critical point in both the
symmetric and broken phases, and correlation lengths of order $10$ are
measured.
The critical indices $\beta_{mag}$ and $\delta$ also pick out mean field
behavior.
We trace the breakdown of the standard scenario (dimensional
reduction and universality) to the composite character of the mesons
in the model. We point out that our results should be
generic for theories with dynamical symmetry breaking, such as Quantum
Chromodynamics.

We also simulated the $O(2)$ model on $8 \times 16^3$ lattices to establish
that
our methods give the results of dimensional reduction in purely bosonic cases
where its theoretical basis is firm. We also show that $Z_2$ Nambu Jona-Lasinio
models
simulated on $8 \times 16^3$ lattices give mean field rather than three
dimensional
Ising model indices.


\vfill
\eject
\newpage
\setlength{\baselineskip}{23pt}
\pagestyle{plain}
\renewcommand{\thefootnote}{\arabic{footnote}}
\setcounter{footnote}{0}
\setcounter{page}{1}

\section{Introduction}

Considerable work has been done on the physics of the finite temperature
chiral transition in $QCD$ and related
models\cite{Kar_94}\cite{GGP_94}\cite{HK_95}. Although, there is
little disagreement about
the existence and order of the transition,
no quantitative work, simulations or otherwise, has been done
that decisively determines the
univerality class of the $QCD$ transition with two flavors.
At present, lattice simulations
can not distinguish between $O(2)$ and $O(4)$ fits of the
universal features of the transition\cite{Kar_94}. In fact, they can
not even rule out any $O(N)$ fit decisively. Universality arguments
\cite{PW_83}
have been invoked in
order to reduce the problem to a manageable form, and the analysis of
lattice data has used these frameworks. Until recently, universality
arguments were the most appealing ones available, and due to their beauty and
simplicity were by and large accepted as true. In
essence, they can be phrased as follows.
At finite temperature phase transitions, the correlation length diverges
in the transition region and low energy physics is
expected to be dominated
by soft modes. The contribution of non-zero modes should be suppressed.
Consequently, fermions, which satisfy antiperiodic boundary conditions,
and do not have zero modes, are expected to
decouple from the scalar sector, together with the other non-zero modes.
The theory effectively reduces to a lower dimensional
scalar field theory. This is an important point since it renders the
treatment of such complicated theories as $QCD$ to that of the
three dimensional $\sigma$ model. This argument is based on the universality
hypothesis and it is hard to see how it can fail.

In ref.\cite{KK_95} we presented analytic and numerical evidence
that this scenario does not hold in the Gross-Neveu model.
We argued that dimensional reduction does not necessarily
hold when chiral symmetry breaking is accompanied by composite
mesons. We analyzed the three dimensional Gross-Neveu model at finite
temperature where the global symmetry is $Z_2$ and universality arguments
imply that the finite temperature restoration transition should be
in the universality class of the two-dimensional Ising model. Using the
large-$N$ expansion and computer simulations, we found no hint of such
behavior, but observed mean-field scaling. Several aspects of
the analysis presented in \cite{KK_95} are, however, subject to criticism.
We address those
criticisms in this paper and present new analyses and simulations.
The criticisms are:

1. {\it The large-$N$ analysis could be misleading}.
After all, this happens in a rather dramatic way in the two-dimensional
Gross-Neveu model and could be happening above two dimensions as well
\cite{DMR_75}.
Thus, it is conceivable that the leading order expansion gives
incorrect exponents and the correct ones are recovered only when $N$ is finite.
Our simulations, presented in \cite{KK_95},
were performed for $N=12$ species, and
the coincidence of the large-$N$ and lattice results might be caused by
the fact that the data are not sufficiently accurate to see the $1/N$
corrections. This is certainly possible and, for that reason, we studied
and simulated
a low-$N$ theory, namely $N=4$. We found no change in our conclusions --
the scaling remains mean-field with a high degree of confidence.
Large-$N$ results are summarized in Sec. 4. $N=12$ simulations are given in
Sec. 6. The results for $N=4$ are given in Sec. 7.

2. {\it The scaling region could be too narrow at large-$N$}. One might
argue that
the true scaling behavior is not
mean-field as we observe, but is undetected in our simulations because the
scaling
region is inaccessible to the lattices that we work on. In other words, larger
lattices and larger correlation lengths are needed to
see the two dimensional (2d) Ising scaling predicted by dimensional reduction
in the three dimensional (3d) case. There are two aspects of this objection.
First, for the Gross-Neveu model at zero temperature, Landau
behavior is not marked by
mean-field exponents. Because of the presence of the fermions, there are
long-range interactions that change the universality class and the scaling
is different from local bosonic models with the same symmetry. This
statement is uncontroversial since it is well known that
the scaling laws of the
Gross-Neveu model\cite{HKK_91},
in any dimension below four, have nothing to do with the Ising
model. The discussion of this problem is presented in Sec. 5.
Thus, the 'canonical' belief that the scaling dynamics and widths of
critical regions here is the same as that of BCS
superconductors is not true
 -- the mean-field scaling observed at finite temperature is not Landau
behavior that should be followed by non-Gaussian behavior once the true scaling
region is reached.

Nevertheless, to clarify these points
further, we first wanted to see if there might exist some intrinsic
difficulties
in observing dimensional reduction in lattice simulations. For that
purpose, we simulated a scalar $O(2)$ model which we know
undergoes dimensional reduction and we found an unambiguous and clear
confirmation for its presence already on lattices of modest sizes,
considerably smaller than we used for the Gross-Neveu model. We present these
findings in Sec. 3.

3. {\it Is it possible to imagine a physical picture that
would be consistent with the failure of dimensional reduction?}
When the relevant length scale diverges,
the dimensional reduction scenerio \cite{PW_83} states that the non-zero modes
should decouple and the resulting theory can be described in terms of
light, pointlike mesons. Thus, a simple, familiar physical picture accompanies
the dimensional reduction scenerio. In Sec. 2. we show
that our scenerio, where dimensional reduction does not occur,
is also accompanied by a consistent
physical picture. The loophole to the 'standard' scenerio
in the case of fermions is that the mesons
are composite and that, because of chiral symmetry, their size scales as the
correlation length, e.g. $R_\pi \sim 1/f_\pi$, and diverges as the critical
point is approached. Thus, the meson size can not be neglected on
physical length scales and the fermionic substructure is always apparent. This
type of behavior is nothing new -- it has been observed and commented on
in the context of quantum mechanics and model field theories \cite{JM_92} many
times in the literature. In other
words, the effective theory near the chiral phase transition can not be
described in terms of locally interacting bosonic variables only.
We also point out some physical
differences between the chiral restoration and superconducting transitions
at finite temperature. For superconductors the size of a Cooper pair
increases as the critical temperature is approached. Unlike the Gross-Neveu
model,
where fermion pair condensation occurs as well, in superconductors, the
range of interaction is determined by the size of Cooper pairs. Therefore,
increasing the size of a pair promotes the importance of
long-range interactions
and, as a consequence, the theory's scaling laws are rather well described
by mean field theory. This occurs
over a wide
range of temperatures, even very close to the critical point.
This is the
reason for having a narrow scaling region at the superconducting transition --
the long range interactions promote mean field behavior and only extremely
close to the critical point are the critical fluctuations, which are expected
in an interacting three dimensional system, sufficiently strong to be
detected experimentally. As the critical point is approached to extraordinary
accuracy
in BCS superconductors, the increase in the size of
Cooper pairs leads to a regime where the size of pairs is too large to support
superconductivity --
the relevant physics occurs on the scale where pairing is not visible and
the system returns abruptly to the normal state. The transition to the
normal state is, in fact, weakly first order, although, due to the long range
interactions,  the scaling appears to be well
approximated by mean field theory over a wide range of temperatures.
In the Gross-Neveu model\cite{GN_74} the situation is completely different:
the range of interaction does not scale like the
length scale associated with the condensate -- long
range interactions exist already at zero temperature and
are essential for its non-Gaussian
behavior. Without the long range interactions, the scaling
in the zero temperature large-$N$ Gross-Neveu model
would follow canonical mean field theory.
Thus, although the pairs' size diverges as the critical
point is approached, it is the screening of the long range interactions
caused by finite temperature effects that is responsible for the emergence
of mean field behavior at the chiral restoration transition of the Gross-Neveu
model.

Our main results concerning critical exponents of the finite temperature,
chiral
symmetry restoration transition in Gross-Neveu models read:
$\beta_{mag} = 1/2, \delta = 3,
\gamma =1$. This should be contrasted with the 2d Ising values
$\beta_{mag} = 1/8, \delta = 15,\gamma =7/4$ that, according to the dimensional
reduction scenerio, should describe the finite temperature transition of
the 3d $\sigma$ model.
Note that the most dramatic difference
is between $\delta$'s -- they differ by a factor of 5 and it is hard to
imagine that the lattice simulations could mistake one for the other.
In addition, we measured the universal amplitude ratios\cite{ID_89}.
The quantity that we
found useful in this context is the susceptibility
amplitude ratio $C_+/C_-$ defined
by $\chi_{\pm}^{-1} = C_{\pm} |t|^\gamma$, where $C_+$ and $C_-$ refer to
the broken and symmetric phases, respectively. The ratio $C_+/C_-$ is
universal. In mean field theory $C_+/C_-= 2$, while for the $2d$
Ising model $C_+/C_-= 37.69$. This quantity is especially
suitable to test for
the absence of dimensional reduction since the amplitude ratios in the two
scenerios
differ by nearly a factor of 20, and
the chance for getting an ambiguous answer here is very small.
Our simulations  for the amplitude ratio are very close to the mean field value
and any attempt to accomodate higher values failed. It will also prove valuable
to estimate the correlation length $\xi$ in our simulations. It diverges
at the critical point with the exponent $\nu$, $\xi^{-1} = D_{\pm} |t|^\nu$,
where $D_+$ and $D_-$ refer to the broken and symmetric phases respectively.
The ratio $D_+/D_-$ is also
universal. In mean field theory $D_+/D_-= 2^{1/2}$, while for the $2d$
Ising model $D_+/D_-= 2$. Since we will have explicit calculations of the
correlation length in the broken phase, these universal relations will
allow us to calculate the correlation length in the symmetric phase in
both scenerios.
\vskip10truemm

\section{Physical picture}

The main appeal of dimensional reduction is the simplification it
introduces by assuming that the all non-zero modes decouple. This is the
source of major simplifications since the fermions, being antiperiodic,
should completely disappear from the low energy physics. Where can this
reasoning fail? First, one should
be aware that universality arguments do not
hold for systems with long range interactions. This is especially relevant
for the Gross-Neveu model where long range interactions
play a crucial part. Nevertheless, it
is not necessary to invoke renormalization group arguments in order to see that
fermions should not decouple at the chiral transition. The argument can
be made much simpler using quantum mechanics only. For that purpose, we
digress in this section and consider some simple quantum mechanical models
that illustrate the role fermi statistics play in situations where the
overlap of wave-functions becomes substantial. The aim of this
section is to show that decoupling of the fermions, in fact, is {\it not}
consistent with the physics of chiral symmetry restoration.
In the vicinity of the phase transition,
compositeness of the light mesons has some interesting implications for
low-energy physics.

As an example, we
consider a quantum mechanical system of relativistic
fermions in a box. This model has been taken from the context
of nuclear physics, and
adapted for our purposes with minimum modifications\cite{Lip}.
All the arguments present
standard material in many-body physics and we merely paraphrase them below.
We construct the two-body mesonic states as,

\be
\label{1.1}
|p>\equiv C^\dagger (p)|0>= \int_k f_p(k) b^\dagger(k+p)d^\dagger (k)|0>
\ee
where $b^\dagger , d^\dagger$ are fermion and antifermion creation operators
respectively. The spin indices have been suppressed for simplicity.
The wave function, $f_p(k)$, satisfies the normalization condition
$\int_k |f_p(k)|^2 =1$. Using the anticomuting properties of fermions,
we obtain the expressions for the
canonical comutation relations for the two body operators,

\be
\label{1.2}
[C(p), C(p')]=0,\,\,\,\,\,\,\,\,
[C(p), C^\dagger (p')]=\delta_{p,p'} - \Delta(p,p')
\ee
In general, $\Delta \not= 0$. It is a measure of the deviation from the
purely bosonic behavior of the mesons.
For $p=p'$, $\Delta(p,p)$ has a simple physical
interpretation.

\be
\label{1.3}
\Delta(p,p)=\int_k |f_p(k)|^2 (n(k+p)+\bar n(k))
\ee
where $n(p), \bar n(p)$ are number operators for fermions and antifermions.
The operator $n(k+p)+\bar n(k)$ counts the number of fermions plus
number of antifermions with momentum difference $p$ and is equal to the number
of mesons with momentum $p$.
The weight factor normalizes it to the total number of states in the box.
Therefore, $\Delta$ is the ratio of the number of mesons
to the total number of states in the
system. If the number of composite
states is small, $\Delta \approx 0$ and the mesons
behave approximately as bosons.
Their structure can be neglected and there is little error in
describing the system as a gas of pointlike bosons.
If, on the other hand, the meson density is high, i.e. the number of
meson states is comparable to the total number of states, there is a large
number of overlapping fermions and the composite structure of the mesons is
revealed. This is reflected in the nonvanishing value of $\Delta$.
The bosonic description is inadequate in this regime.

As seen by this example, even if the theory possesses no fundamental
pointlike fermion
excitations ( like $QCD$, for example), the fermionic nature of its
constituents gives rise to two regimes:
1) when the density of composite mesons is
low, the theory is well described by an equivalent bosonic model;
2) when the mesons are dense, the bosonic description fails;
the fermions become essential degrees of freedom irrespective of how heavy
they are.

To make a connection with the chiral restoration transition, we
recall the effect of finite temperatures on fermions and mesons. First,
since they are the lightest particles in the system,
composite mesons are easily
excited by thermal effects. Fermions, although light in the critical
region, do not appear so in the Euclidean theory. Due to antiperiodic
boundary conditions, their Matsubara frequencies are odd
and they do not possess a zero mode.
Thus, on the energy scale of the scalar sector, they appear as heavy and
might be expected to decouple at low energies.  The zero mode component
of the scalar sector would then give
the dominant contribution in the scaling region and the
effective theory would be a dimensionally reduced scalar model.
This is the reasoning behind the standard scenario\cite{PW_83}.

The problem that we are addressing in this paper concerns
the adequacy of the
effective scalar theory. In terms of the meson degrees of freedom the
applicability of the effective scalar theory can be
phrased in terms of how dense the
mesonic gas is near the critical temperature. There are two
possibilities. The first one corresponds to the standard scenario where the
meson size does not change appreciably as the
temprerature increases from zero. In this case in the critical
regime where the thermal correlation length is large,
the system can be treated as a dilute mesonic gas and
an effective scalar theory could apply.
In the alternative realization, the
meson size increases significantly near the critical temperature.
If this is the case, then the effective scalar theory has to break down
even in the broken phase.

We briefly examine which of the two alternatives is likely to occur in
the case of the chiral transition.
Heating the system causes an increase in the thermal motion of the
particles. As the temperature increases,
the contituent fermions inside a meson
will wiggle at an increasing rate. One expects that the total
energy of the composite will follow the increase in the zero-point energy
of constituents. In the phase without spontaneous symmetry breaking,
this results in a well known linear increase of the meson masses
with temperature. How can this be reversed in
the broken phase where the symmetry restoration
requires that the meson masses gradually decrease to zero as
the system is heated? The only way to
neutralize an increase in zero point energy due to a temperature increase
is to allow the radius of the meson to increase as well.
Any mechanism that would prevent this from happening would increase
the energy of the system. Furthermore,
the size of composites should diverge as
the correlation length, since if it didn't, the energy of the
system would not scale properly -- introducing a composite of
finite extent into the critical system would increase the total energy by a
finite amount that does not vanish at the critical point.

This being the case, the transition region can be described
as a system of highly overlapping composites. In this regime, the violation
of the bosonic character of the mesons is maximal.
The scale on which quantum fluctuations that cause the binding to occur
is the same as the thermal correlation length. Thus, the classical description
in terms of an effective scalar model is not adequate in this regime.
The effective scalar theory
fails to capture the relevant degrees of freedom even in the broken
phase, and this might, in principle, occur well before the critical
temperature is reached.

These arguments do not depend on whether the fermions
are confined or not. The effect of having fermion constituents, confined or
not, implies that by populating the system with composite
mesons certain energy levels become inaccessible. As a consequence, at
low energies the effective mesonic
interaction is less operative and the binding energy
is reduced. Thus, weaker binding occurs at higher
densities. Consequently, the mesons are fluffier as their density
increases. Allowing fermions to be thermally excited as well, as
happens in the absence of confinement, does not alter this picture
qualitatively. On the contrary, their presence
combines constructively with Pauli blocking caused by the mesons.
\vskip1truecm

\section {  Dimensional reduction in a scalar model}

\noindent --{\it $O(N)$ Scalar Theory}

To illustrate the idea of dimensional reduction and its relevance for the
phase transitions at finite temperature,
we start with the simplest possible example of an
$N$-component scalar theory and consider
the large-$N$ limit\cite{DJ_74} for simplicity.
The potential of the model has the
standard form:
$V(\phi )=1/2 \mu^2{\vec\phi}^2 +\lambda/4 ({\vec\phi}^2)^2$.
Complications due to Goldstone bosons can be avoided by
by working in the symmetric phase.
To leading order in $1/N$, the corrections
to the propagator are given by the single tadpole contribution.
Since the tadpole is
momentum independent, it effects only the susceptibility and not the
wave function renormalization.
The susceptibility, $\chi =\int_x <\phi(x)\phi(0)>_c$, is the zero-momentum
projection of the correlation function and at zero temperature it is given by

\be
\label{s.1a}
\chi^{-1}=\mu^2 +\lambda\int_q {1\over{q^2+\chi^{-1}} }
\ee
where $\int_q = \int d^dq/(2\pi )^d$, and
we absorb the combinatorial factor into $\lambda$.
Defining the critical curvature $\mu^2_c$ as the point where the
susceptibility diverges ($\mu_c^2+\lambda\int_q 1/q^2=0$), the
expression for the inverse susceptibility can be recast into

\be
\label{s.1b}
\chi^{-1}\Biggl( 1+\lambda\int_q {1\over{q^2(q^2+\chi^{-1})} }\Biggr)
=\mu^2-\mu_c^2
\ee
The extraction of the critical index $\gamma$ reduces to counting powers
of the infrared (IR) singularities on the left hand side (LHS) of
eq.(\ref{s.1b}).
Above four dimensions, both terms are IR finite and the
scaling is mean field ($\gamma =1$). This is supplemented by
logarithmic corrections in four dimensions.
Below four dimensions the second term in eq.(\ref{s.1b}) dominates
the scaling region -- the integral diverges as $\chi^{(4-d)/2}$.
This gives the zero-temperature susceptibility exponent $\gamma=2/(d-2)$
\cite{DJ_74}.

The finite temperature susceptibility is obtained from eq.(\ref{s.1a}) after
replacing the frequency integral with the Matsubara
sum. For a given value of $\mu^2$
we define the critical temperature, $T_c$, by

\be
\label{s.2}
\mu^2 +\lambda T_c\sum_n\int_{\vec q} 1/(\omega_{nc}^2+{\vec q}^2)=0
\ee
where $\omega_{nc}=2\pi nT_c$.
The momentum integrals are now performed over $d-1$ dimensional space.
This equation defines the critical surface in the space spanned by
$\mu^2$, $\lambda$, and $T_c$ which reads

\be
\label{s.3}
\mu^2-\mu_c^2 +a\lambda T_c^{d-2}=0
\ee
where $a$ is some positive constant. For fixed value of the curvature,
$\lambda$ decreases as the temperature grows. Conversely, for a fixed value
of the coupling, the temperature tends to flatten the curvature. This effect
is controlled by the strength of the coupling and is reduced at small
$\lambda$. One would expect that perturbation theory becomes
progressively more reliable as the system is heated.
Inserting eq.(\ref{s.2}) into
the expression for the inverse susceptibility results in

\be
\chi^{-1}\Biggl( 1+\lambda T_c\sum_n\int_{\vec q}
{1\over{(\omega^2_{nc}+{\vec q}^2)(\omega^2_n +{\vec q}^2+\chi^{-1})} }\Biggr)
\ee
\be
\label{s.4}
=\lambda T_c\sum_n\int_{\vec q}
{{{\vec q}^2(T/T_c-1)+\omega_n^2(T_c/T-1)}
\over{(\omega^2_{nc}+{\vec q}^2)(\omega^2_n +{\vec q}^2+\chi^{-1})} }
\ee
Separating the $n=0$ mode $(\omega_0 =0)$
from the rest of the sum, we get the leading singular behavior

\be
\label{s.5}
\chi^{-1}\Biggl( 1+\lambda T_c \int_{\vec q}
{ 1\over{{\vec q}^2({\vec q}^2+\chi^{-1})} }+\sum_{n\not= 0}...
\Biggr)=
\lambda T_c\int_{\vec q}{  { T/T_c -1}\over{{\vec q}^2+\chi^{-1}} }
+\sum_{n\not= 0} ...
\ee
The $n=0$ piece dominates the scaling region. It
resembles the zero-temperature expression,
eq.(\ref{s.1b}), except that now,
the integrals are performed in $d-1$ dimensions, instead of $d$. The power
counting is the same as before and it yields the thermal exponent
$\gamma_T=2/(d-3)$. This is the same as the zero-temperature $\gamma$ in
$d-1$ dimensions \cite{DJ_74}. Other
critical exponents show the same type of behavior as $\gamma$.

Neglecting the non-zero modes, which is justifiable in the scaling region,
the expression in eq.(\ref{s.5}) resembles that of the $d-1$ theory with
effective coupling $\lambda_{eff}=\lambda T_c$. This is in the spirit of
dimensional regularization since the dimensions of quartic couplings in scalar
theories in $d$ and $d-1$ dimensions differ by one unit. As long as $d>3$,
the system is perturbative around the critical line -- the effective coupling
$\lambda T$ is small compared to $\lambda T\sim  T_c^{3-d}$  even at high
temperature.

This large-$N$ exercise illustrates the mechanism of
dimensional reduction and was chosen here for its
simplicity. A technically more complicated
case of finite-$N$ models can be carried out using the
$\epsilon$-expansion, for example. The arguments presented above
are based on counting only the light degrees of freedom and are thus generic.
It would be difficult to imagine how
they can fail in the context of scalar models. We undertake a
numerical study of the finite temperature
transition in the four-dimensional $O(2)$ model, believing that dimensional
reduction occurs and consider under what conditions and degree of confidence
we can establish it numerically.
This exercise is performed for several reasons. Before going on with
the study of fermionic models, we want to determine
some criteria for establishing dimensional reduction numerically by simulating
the model when the results are known. For that purpose we choose to work
with the simplest scalar theory. In particular, we want to know the
lattice sizes required, the statistics needed etc..
\vskip5truemm

\noindent
--{\it Computer simulations of the $O(2)$ $\sigma$-model}

We simulated a $8 \times 16^3$ lattice using the O(2) model code we used
in a recent study of scalar electrodynamics \cite{BK_93}. The code is a
relatively
conservative one which combines the Metropolis and over-relaxation
methods. It is discussed in greater detail in ref. 13. It was adequate for our
exploratory studies here, but will be replaced by a cluster algorithm in
future, more accurate studies. As we shall see, the evidence for
dimensional reduction was very clear, so it served our purposes.
The Euclidean action of the model reads,

\be
\label{s.6}
S=-\beta\sum_{x,\mu}(\phi_{x}^{\ast} \phi_{x+\mu}+c.c.)
\ee
where $\beta$ is the coupling constant and $\phi_{x}=exp(i\alpha(x))$
is a phase factor at
each site of the lattice. At each
coupling $\beta$, we typically ran $10^6$ sweeps of the algorithm and
measured the mean magnetization $\sigma$ and its susceptibility $\chi$.
These measurements were first done without any external magnetic field
in the system because we want to obtain the critical coupling $\beta_c$
and the critical index $\beta_{mag}$ for the magnetization transition.
Since the mean magnetization would average to zero in such a computer
experiment without a symmetry-breaking field, we actually measured the
magnitude of the magnetization and used this as our 'effective' order
parameter. This is a standard procedure in such cases, and is reliable
in the broken symmetry phase where the mean magnetization is substantial.
It is not correct too close to the critical point and in the symmetric
phase. If one were simulating the Ising model, for example, then vacuum
tunneling on the finite lattice would restore its global $Z_2$ symmetry. These
tunneling effects diminish on larger lattices and they determine the
range of couplings that can be reliably simulated on a fixed lattice size.
Our data for the O(2) model on the $8 \times 16^3$ lattice is accumulated
in Table 1. The error bars recorded there account for the statistical
uncertainties in each $10^6$ sweep run and were estimated by the usual binning
procedures which account for the correlations in the raw data sets. We
found that we could not reliably measure the order parameter of
the model for couplings below .1525 due
to tunneling effects. The magnetization data was fit to the form
$\sigma = A(\beta-\beta_c)^{\beta_{mag}}$ for couplings between .17 and .155.
The data and the fit are shown in Fig.1. The simple powerlaw fit is very good
and results in the predictions $A = 1.97(7)$, $\beta_c = .1519(1)$ and
$\beta_{mag} = .34(2)$. The confidence level for the fit is .50.
The best measurements of the critical indices of
the three dimensional O(2) model yield $\beta_{mag} = .346(2)$. So, our
relatively modest simulation is in good agreement with the
dimensional reduction hypothesis and certainly distinguishes mean field
exponents where $\beta_{mag} = .50$.

Next we consider the susceptibility which should diverge in the critical
region with a powerlaw singularity controlled by the index $\gamma$.
On a finite lattice it is conventional to parametrize such data with the form
$1/\chi = A(\beta-\beta_c)^{\gamma} + C$.
The data in Table 1 fit this
simple powerlaw hypothesis over the coupling range .17 to .155
(as shown in Fig.2) and a fit
with confidence level .41 predicted the parameters $A = 71.5(3.7)$,
$C = .029(2)$ and $\gamma = 1.22(1)$ for the choice $\beta_c = .1519$
found in the magnetization data. We are particularly
interested in the universal
index $\gamma$ here and we found that it was not sensitive to the precise
value of $\beta_c$ chosen in the fitting form. Our predicted value for
$\gamma$ is close to the best estimate of $\gamma$ for the three dimensional
O(2) model, $\gamma = 1.32(2)$, and is rather far from the mean field
presiction of $\gamma = 1.00$. The three dimensional O(2) model result
of $\gamma = 1.32(2)$ is based on a thorough finite size scaling analysis
and its uncertainty reflects that, unlike our relatively modest calculation
which extracts an estimate for $\gamma$ from one lattice size and just records
the uncertainty in a single fit. A thorough finite size analysis is
planned, but our purpose here is just to indicate that a single
simulation on a $8 \times 16^3$ lattice gives good evidence for the
dimensional reduction scenerio. In particular, we find no need for
enormous lattices. In addition, the critical region appears to be
easily accessible and naive powerlaw fits work well. Although we did
not measure the system's correlation length (inverse of the scalar
mass) with enough precision to extract the critical index $\nu$, we
did confirm that the correlation length is just a few lattice spacings
throughout the region of parameter space where we extracted the critical
indices of the finite temperature transition. It was {\it not} necessary
to work so close to the transition itself that the correlation length
would be large compared to the temporal extent of the lattice. In other
words, we confirmed the well-known result in statistical mechanics and
computer simulation physics, that once the correlation length is comparable to
the temporal extent, then the finite temperature critical indices control
the simulation data. The importance of this comment lies in the fact that
formal derivations of the dimensional reduction scenerio require that
the zero mode dominate the Matsubara sum and this occurs when the scalar
mass is small compared to the temperature. Numerical simulations suggest
that the result of the dimensional reduction scenerio -- that the critical
indices of the finite temperature transition are those of the three
dimensional O(2) model -- is more robust than its formal derivation. Apparently
the higher Matsubara frequencies that are simply ignored in the formal
derivations do not effect the critical singularities but can be
absorbed into noncritical, nonuniversal aspects of the transition. We suspect
that this fact applies to other models besides those studied here.

Next we turned to our most discerning measurement - the index $\delta$. Recall
that at the critical point $\beta_c$, the mean magnetization should depend
on an external symmetry breaking field m as $\sigma = Am^{1/\delta}$. Here
we are using the notation 'm' where magnetic field 'h' would be more
conventional. We do this because we will be turning to chiral symmetry
below and there the symmetry breaking field is m, the bare mass of the
fermion. We simulated the O(2) model on the $8 \times 16^3$ lattice at
our best estimate of $\beta_c = .1519$. We considered symmetry breaking
fields m ranging from .002 to .05 as shown in Table 2. Again, $10^6$
sweeps of the algorithm were done at each parameter setting. On a modest
$8 \times 16^3$ lattice, we were not able to sensibly simulate smaller
m values due to vacuum tunneling. Fits to the data over the m range of
.002 - .010 were very good, producing confidence levels of .92 and the
parameters $A = .773(2)$ and $\delta = 4.7(2)$. The fit is shown in Fig.3
and 4 as $\sigma$ vs. $m$ and $ln(\sigma)$ vs. $ln(m)$. The value of $\delta$
for the three dimensional O(2) model is $\delta = 4.808(7)$, so we have
further, quite decisive, evidence for the dimensional reduction
scenerio. The mean field value of $\delta = 3.00$ is easily rejected.

In summary, conventional simulations of the O(2) model on a $8 \times 16^3$
lattice are in good agreement with the dimensional reduction scenerio
expected of a purely bosonic theory. Powerlaw fits work well and the
critical region, as parametrized in conventional fashion, is accessible
even on a lattice of modest size and asymmetry. The critical indices
predicted by dimensional reduction are found over a region of couplings
where the correlation length is just several lattice spacings. Huge lattices
and huge correlation lengths exceeding the temporal extent of the lattice
are not necessary when measuring the critical indices of interest to the
accuracy discussed in this article.
\vskip1truecm

\section{ Large-N Analysis of the Gross-Neveu Model at Finite Temperature}

Large-$N$ studies of the four-fermi models proved successful in understanding
their fixed point structure and nonperturbative renormalization.
At zero temperature the critical coupling that separates the two phases
is identified as an ultra-violet (UV) fixed point around which an interacting
continuum limit can be constructed. The emerging critical
exponents are non-Gaussian and obey hyperscaling \cite{HKK_91}.
The validity of the large-$N$ expansion
has been shown to hold using explicit expansions up to $1/N^2$ \cite{G91}.
These
studies find strong support in
lattice simulations of the model with both $N=2$ \cite{KLLP_94} and
$N=12$ \cite{HKK_93}. Recently, a rigorous
proof in 3d has been constructed.
Thermodynamics of the Gross-Neveu model has been studied extensively and
the results of the leading order calculations both in three and four
dimensions produced mean field exponents \cite{RWP_89}.
These results are in conflict
with dimensional reduction \cite{PW_83}
since the model possesses  $Z_2$ symmetry
and its thermodynamics would be expected to lie in the universality
class of a dimensionally reduced
Ising model. However, this result should not be that surprising since at
zero temperature the symmetry is Ising as well, but the
critical exponents of Gross-Neveu model are far from the Ising values.
This discrepancy
can be attributed to the presence
of massless fermions in the model which induce
long range interactions that cause violations
of universality. One could argue, on the other hand, that at finite temperature
long range interactions are absent due to Debye screening and
fermions receive, even in the symmetric phase,
thermal (chirally invariant) masses equal to their lowest Matsubara frequency
$\omega_0=\pi T$.

This apparent departure of the large-$N$ result from the expected
scenario of dimensional reduction remained by and large unnoticed by the
theory community chiefly because of the belief that
large-$N$ expansions are misleading at leading
order and that corrections were needed to get the
correct physics. This is well known to be the case in two dimensions where
the failure of the large-$N$ limit is most dramatic --
not only does it predict erroneously a finite, instead of zero,
restoration temperature, but it also fails to capture the dynamics of
kinks and their condensation.
In ref.\cite{HKK_93}
we carried out numerical simulations of the 3d Gross-Neveu model
with the aim of verifying or refuting the predictions of the large-$N$
expansion.
We pointed out
that, in fact, above two dimensions, the large-$N$ expansion can be
trusted and that the
existing results seem to provide a good guide to the
thermodynamics of the model.
Results of the lattice simulations that $\delta=3$ and $\beta_{mag}=1/2$
seem robust because dimensional reduction predicts 15 and 1/8 for the
two exponents, respectively.

This controversial result comes very naturally in the large-$N$
expansion and is attributed to the compositeness of the scalar mesons.
However, some of its aspects like the precise values of the exponents,
their sensitivity to the change in dimensionality, the width of
the critical region and, therefore, the credibility of the data can be
subjected to greater scrutiny.
In this section, we extend the analysis of the thermodynamics of the
three dimensional
Gross-Neveu model outlined in ref.\cite{HKK_93} and extend it to
the four dimensional model as well. One of the reasons for doing the
four dimensional model is to see what ingredients are relevant for
the failure of dimensional reduction. In particular, the difference between
three and four dimensional models is that the two have different relevant
operators. For example, at the fixed point
where the dimension of the $\sigma$-field is 1 independent of $d$,
$\sigma^4$ in four dimensions is marginal, while in
3$d$ it is irrelevant. In four dimensions, therefore, there is a
quartic term and dimensional reduction can not be ruled out even
if one accepts that it does not occur in the three dimensional model.
Also, it is of interest to compare the scaling regions and their widths
in the $O(2)$ and Gross-Neveu models and confront the fermion model
with data of comparable (or better) quality than that taken in our
numerical study of the scalar theory where
dimensional reduction was established.

As we announced in the introduction, we expect the departure from dimensional
reduction in the fermionic models on physical grounds. Our arguments are
based on the expectation that the effective description provided by the bosonic
theory
fails in the vicinity of the restoration transition since the size of
the composite mesons scales as the thermal correlation length and
at low-energies scalars are not the only relevant degrees of freedom.
This scaling of
the meson size, we argued in the introduction, is implied by
the symmetry restoration nature of the transition.
To put things into perspective, we
recall that according to standard arguments, fermions are expected to
decouple near the phase transition and the effective theory is described
in terms of mesonic degrees of freedom. The reason for fermionic decoupling
is that their Matsubara frequencies are odd integer multiples of the
temperature, due to their antiperiodic boundary conditions.
Their zero modes are missing, and from
the point of view of the dimensionally reduced scalar sector, fermions
appear as particles with masses of the order of $T$; they are
much heavier than the scalars and decouple. In what follows we
discuss in some detail why this does {\it not} happen.

We proceed along the same lines as in the scalar theory.
We analyze the problem of chiral symmetry
restoration in a Gross-Neveu model given by the lagrangian
$L=\bar\psi (i\partial +m +g\sigma )\psi -
{1\over 2}\sigma^2$.
When fermions are integrated out of the model, the Ising symmetry,
$\sigma\to -\sigma$, of the effective action  becomes manifest.
First, we start with the zero-temperature gap equation and
corresponding critical exponents.
The model can be treated in the
large-$N$ limit. To leading order, the fermion self-energy, $\Sigma$,
comes from the $\sigma$-tadpole: $\Sigma=m-g^2<\bar\psi\psi>$.
To obtain the scaling properties of the theory, we define
the critical
coupling as $1=4g_c^2\int_q 1/q^2$. Combining this definition with the gap
equation leads to

\be
\label{f1}
{m\over\Sigma}+\bigl( g^2/g_c^2 -1\bigr)=
4g^2\int_q {{\Sigma^2}\over{q^2(q^2+\Sigma^2)}}
\ee
Like the scalar example,
this form is especially well suited for extracting critical indices
since the problem reduces again to the
counting of the infra-red divergences on the
right hand side \cite{Z-J_91}\cite{HKK_91}. The critical indices are defined by
$<\bar\psi\psi>|_{m=0}\sim t^{\beta_{mag}},
<\bar\psi\psi>|_{t=0}\sim m^{1/\delta},
\Sigma |_{m=0}\sim t^\nu,$ etc.. Here,
$t=g^2/g_c^2 -1$ is the deviation from the critical coupling.
Since $\Sigma\sim <\bar\psi\psi>$, $\beta_{mag}=\nu$ to leading order.
Above four dimensions the integral in  eq.(\ref{f1}) is finite
in the limit of vanishing $\Sigma$ and the scaling is mean-field.

Below four dimensions, the $\Sigma\to 0$
limit is singular -- the integral scales as
$\Sigma^{d-2}$.
Thus, in the chiral limit $t\sim \Sigma^{d-2}$, and at
the critical point, $t=0$,
away from the chiral limit, $m\sim\Sigma^{d-1}$.
The resulting exponents are non-gaussian: $\beta_{mag}=1/(d-2)$ and
$\delta = d-1$. The remaining exponents are obtained
easily: $\eta=4-d, \gamma=1$ \cite{HKK_91} and one can check that they
obey hyperscaling.

We now consider the Gross-Neveu model at finite-temperature.
We choose to stay between two and four dimensions
to emphasize how zero-temperature powerlaw scaling
changes at finite temperature. The gap equation is now modified to

\be
\label{f2}
\Sigma=m+4Tg^2\sum_n\int_{\vec q} {\Sigma\over{\omega_n^2 +\vec q^2
+\Sigma^2}}
\ee
where $\omega_n=(2n+1)\pi T$.
For $g>g_c$ the critical temperature is determined by:

\be
\label{f3}
1=4T_c g^2\sum_n\int_{\vec q} 1/(\omega_{nc}^2+{\vec q}^2)
\ee
where $\omega_{nc}=(2n+1)\pi T_c$.
Combining the definition of $T_c$ with the finite-temperature
gap equation, we can
bring it to a form similar to eq.(\ref{f1})

\be
\label{f4}
{m\over\Sigma}=\bigl( 1-T/T_c \bigr)+
4Tg^2\sum_n\int_{\vec q} {{\Sigma^2+\omega_{nc}(\omega_n+\omega_{nc})(T/T_c-1)}
\over{(\omega^2_{nc}+{\vec q}^2)(\omega_n^2+{\vec q}^2+\Sigma^2)}}
\ee
The extraction of the critical exponents proceeds along
the same lines as in the zero-temperature case.  One difference
relative to  eq.(\ref{s.5}) becomes apparent immediately:
the zero modes are absent here and the integrand in
eq.(\ref{f4}) is regular in the $\Sigma\to 0$ limit even below four dimensions.
Consequently, the IR divergences are absent from all the integrals and
the scaling properties are those of mean-field theory:
$\beta_{mag}=\nu=1/2, \delta=3$, etc.
This is true for any $d$,  below or
above four. It appears that in this case, contrary to the scalar example,
the effect of making the
temporal direction finite ($1/T$)
is to regulate the IR behavior and
suppress fluctuations. This is
manifest in other thermodynamic quantities as well. For example,
to leading order, the scalar
susceptibility, $\chi =
\partial <\bar\psi\psi>/\partial m$, is given by

\be
\label{f5}
\chi^{-1}=8g^2 T\sum_n\int_{\vec q} { {\Sigma^2}\over{(\omega_n^2+\vec
q^2+\Sigma^2)^2} }
\ee
Once again, because of the absence of the zero mode ($\omega_0 =\pi T$),
the integral in eq.(\ref{f5})
is analytic in $\Sigma$, and the mean field relation
$\chi^{-1}\sim \Sigma^2$ follows. This is equivalent to $\gamma=2\nu=1$.
The explicit calculation of the momentum dependence of the $\sigma$
propagator yields $\eta=0$.

The phase diagram is given by eq.(\ref{f3}) which defines a critical
line in the $(g,T)$ plane.
For every coupling there exists
a critical temperature beyond which the symmetry is restored. Conversely,
for a fixed temperature there is a critical coupling, defined by the above
expression, corresponding to symmetry restoration.
At zero temperature, the symmetry is restored at $g=g_c$. Thus,
$(g=g_c, T=0)$ is the ultra-violet (UV) fixed point.
As the coupling moves away from $g_c$, a higher
restoration temperature results. At infinite coupling the end-point,
$(g=\infty , T=T_c)$, is the IR fixed point. The critical line
connects the UV and IR fixed points dividing the $(g,T)$ plane
into two parts. The equation for the critical line can be brought into a
compact form by combining the expression for $T_c$ with the
definition of the zero-temperature critical coupling.  This results
in: $(g^2/g_c^2 -1) \sim T_c^{d-2}(g)$, i.e. $T_c(g) \sim \Sigma (T=0)$.
In this way, for any value of the coupling, the critical temperature
remains the same in physical units. This tradeoff of the temperature for the
coupling constant can be seen explicitly in the four-fermi model.
In particular, since the lattice simulations are performed at a fixed number
of temporal links, the critical line is approached by varying $g$. The critical
exponents thus obtained will reflect the singularities as a function of
the coupling, not the temperature. However, the tradeoff between the
two is always linear so that $g-g_c \sim T-T_c$. To show this, we start with
the gap equation on the critical line. We denote the points on the critical
line by $(g^*,T^*)$, to distinguish them from the endpoints $g_c$ and $T_c$.
To show that the above statement holds, we compare the gap equation at
finite temperature with the expression for the critical line (in the
chiral limit).

\be
\label{f6}
1=4T^*g^{*2}\sum_n\int_{\vec q} {1\over{\omega_n^{*2} +\vec q^2}},\,\,\,\,\,\,
\,\,\,\,
1=4Tg^{2}\sum_n\int_{\vec q} {1\over{\omega_n^{2} +\vec q^2+\Sigma^2}}
\ee
Since we already have the expression for the scaling in terms of
temperature and at fixed coupling,
we now fix the temperature to some value $T=T^*$ on the critical line and
approach the critical region horizontally. Simple algebra leads to

\be
\label{f7}
{ {g^2-g^{*2}}\over {g^{*2}}}=
4T^*g^{2}\sum_n\int_{\vec q} {{\Sigma^2}\over{(\omega_n^{*2} +\vec q^2)
(\omega_n^{*2} +\vec q^2+\Sigma^2)}}
\ee
This results in the scaling $\Sigma^2 \sim (g^2-g^{*2})$, which, when compared
with the scaling at fixed $g$, gives $|T-T_c| \sim |g^2-g^{*2}|$. Therefore,
the exponents are independent of the direction with which
we approach the critical line.
\vskip1truecm

\section
{ Landau Theory and Ginzburg Criterion in Theories with Long Range
Interactions}

In this section we consider generalizations of the Landau theory and
Ginzburg criterion for the case of long-range forces.
The particular motivation to analyze this problem in a
separate section is the problem of universality that, in the context
of chiral transitions, requires some attention.
The Gross-Neveu model with interaction term
$L=g^2\sum_i(\bar\psi_i\psi_i )^2$ has discrete chiral symmetry
irrespective of the number of fermion species which is manifest
after introducing an auxiliary field, $\phi$ and
integrating out the fermions. In order to conform with the standard notation,
in this section we use $\phi$ (instead of $\sigma$) to denote the scalar field
and use $\sigma$ to denote the range parameter. The effective action is

\be
\label{g1}
S_{eff}={1\over 2}\int_x\phi^2 -N {\rm tr}\ln (i\partial + m+g\phi)
\ee
Clearly this action has a discrete, Ising-like, symmetry $\phi\to -\phi$.
The unexpected feature of this model which
is at odds with the expectations based on universality is that
the critical exponents are
not Ising. The origin of the failure of the standard universality
arguments is usually attributed to the appearence of massless fermions
that accompany the chiral transition. They are a source of long range
forces which are capable of changing the universality class.
In fact, it was shown in \cite{AK_95} that the
Gross-Neveu model describes the same physics as a Landau theory with
long range forces. However, long range forces alone
can not account for the difference between the Gross-Neveu and Ising
critical exponents. The relationship between fermionic and scalar
$\sigma$-models is more intricate and goes beyond naive universality
arguments. It turns out  that the universality
class of the Gross-Neveu model is that of $O(N=-2)$
magnets with with long range forces \cite{AK_95}.

Since the effective
hamiltonian is not local, we allow for the appearance of
a nonintger power in the Landau free energy. To leading order in $1/N$,
$S_{eff}$ is given by the saddle point approximation. Thus, keeping only
the relevant terms, we obtain from eq.(\ref{g1})

\be
\label{g2}
H=\phi(\nabla^2)^{\sigma/2}\phi
+{1\over 2}t\phi^2 +\lambda \phi^p
\ee
where we leave $p$ undetermined for the time being. The first and third
term in eq.(\ref{g2}) are produced by radiative corrections. The extraction of
the
critical exponents in this case is straightforward. The result is

\be
\label{g3}
\alpha={ {p-4}\over{p-2} },\,\,\,\,
\beta_{mag}={1\over{p-2}},\,\,\,\,
\gamma =1,\,\,\,\,
\delta=p-1,\,\,\,\,
\eta=2-\sigma , \,\,\,\,
\nu={1\over{\sigma}}
\ee
These are Landau exponents obtained by the mean-field treatment of
the Ginzburg-Landau hamiltonian eq.(\ref{g2}).
There are two independent external parameters,
$\sigma$ and $p$, and so hyperscaling is not respected in general.
The value of exponent $\nu$ follows from the expression for the
propagator: $G(k)=1/(k^\sigma +t)$.
Imposing hyperscaling relates $p$ and $\sigma$ by

\be
\label{g4}
p = {{2d}\over{d-\sigma}}
\ee
For this value of $p$, the exponents of eq.(\ref{g2}) read

\be
\label{g5a}
\alpha=2-d/\sigma,\,\,\,\,
\beta_{mag}={{d-\sigma}\over{2\sigma}},\,\,\,\,
\gamma =1,\,\,\,\,
\delta={{d+\sigma}\over{d-\sigma}},\,\,\,\,
\eta=2-\sigma ,\,\,\,\,
\nu={1\over{\sigma}}
\ee
For comparison, we rewrite the Gross-Neveu exponents
\be
\label{g5b}
\alpha={{d-4}\over{d-2}},\,\,\,
\beta_{mag}={1\over{d-2}},\,\,\,
\gamma =1,\,\,\,
\delta=d-1,\,\,\,
\eta=4-d ,\,\,\,
\nu={1\over{d-2}}
\ee
Both sets of exponents are obtained from the stationary point solution which
neglects the fluctuations in the $\phi$ field and, thus, represent
generalized
mean field scaling. It is interesting to compare the two sets of
exponents -- they coincide for
$\sigma=d-2$.
Unlike scalar models with long range forces, where
$\sigma$ is an external parameter, in the Gross-Neveu model $\sigma$
is generated dynamically
by the fermions.
The restriction $2<d<4$ under which the above
exponents were obtained, $\sigma=d-2$ translates into
$0<\sigma<2$. This is exactly the range of values for the $\sigma$ parameter
for which the long range forces are capable of changing the universality
class. This is easy to understand by looking at the gradient term in
eq.(\ref{g2}). Clearly, for $\sigma <2$ the dominant contribution at low
energy comes from the $(-\nabla^2)^{\sigma /2}$ term; the kinetic term,
$(-\nabla^2)$, can be neglected in the IR. The local limit (short range
interaction) is recovered for $\sigma=2$. Note that $\sigma <2$ introduces
anomalous dimensions into the effective theory since the scaling dimension of
the scalar field is determined by its gradient term. From eq.(\ref{g2}), the
dimension of the $\phi$-field is $d_\phi =(d-\sigma)/2$. The absence of
$(-\nabla^2)$ in the effective hamiltonian, eq.(\ref{g2}), can be phrased
also in terms of the compositeness condition, $Z=0$.  In the context of the
Gross-Neveu model this is equivalent
to the statement of the existence of the long range forces.

It is a straightforward exercise to
verify that the equation of state and other universal quantities like
amplitude ratios in the Gross-Neveu model and in Landau theory
coincide after the identification $\sigma=d-2$ is made.
Massless fermions introduce nonlocal effects into the game
and the fermionic model maps onto a $\sigma$-model
with long-range forces. Thus, the universality class of the Gross-Neveu
model is not the standard, short range
Ising model, but a Landau theory with long range
forces and a specific value of the range parameter $\sigma=d-2$.
Thus, the $N=\infty$ limit of the Gross-Neveu model corresponds to a
generalized Landau theory and the exponents of eq.(\ref{g5b})
replace the standard
mean-field ones.

The Ginzburg criterion that determines the importance of the fluctuations, and
therefore the validity of the Landau theory, can be derived for the case of
long-range forces by imposing the following requirement on the second
cumulant

\be
\label{g6}
{\cal E}_{LG}={ {\int_x <\phi(x)\phi(0)>_c}\over{\xi^d <\phi>^2} }\ll 1
\ee
The meaning of this requirement is simple: since the mean field replaces
the value of the field variable with its expectation value, the connected
Green's functions vanish and Landau theory is exact. The Ginzburg criterion,
eq.(\ref{g6}), is a requirement that the actual result is not far from the
gaussian limit. It is expressed as a dimensionless number that measures
a relative error made by the mean-field approximation
assuming the physical volume to be determined by the
correlation length. It relates the bare parameters of the theory ($t$ and
$\lambda$ in this case). From the Landau-Ginzburg hamiltonian, the expressions
for the order parameter and susceptiblity reads

\be
\label{g7}
<\phi>\sim \biggl({ {|t|}\over{\lambda} }\biggr)^{1/(p-2)},
\,\,\,\,\,\,\,\,\,\,
\chi^{-1}\sim |t|,
\,\,\,\,\,\,\,\,\,\,
\xi\sim |t|^{-1/\sigma}
\ee
The Ginzburg criterion, eq.(\ref{g6}), then becomes

\be
\label{g8}
{\cal E}_{LG}={ {|t|^{-1}}\over{|t|^{-d/\sigma}
\biggl({ {|t|}\over{\lambda} }\biggr)^{2/(p-2)}
} }=\lambda^{2/(p-2)} |t|^{(d-d_c)/\sigma}
\ll 1
\ee
where the upper critical dimensionality is defined as $d_c=p\sigma /(p-2)$.
In the Gross-Neveu model $d=d_c$ in the leading order calculations.
The validity of the Landau theory is then conditioned by the requirement
that

\be
\label{g9}
\lambda^{2/(p-2)}\ll |t|^{(d_c-d)/\sigma}
\ee
Clearly, for $d<d_c$ the right hand side of eq.(\ref{g9})
shrinks to zero as the critical point is
approached and mean field theory fails to describe the critical region.
Above $d_c$, the right hand side diverges leaving the coupling $\lambda$
unconstrained; the mean field description in this case is adequate.
The standard Ginzburg criterion for short range forces is recovered
by substituting $\sigma=2$ and $d_c=4$ in eq.(\ref{g9}).

Since the exponents satisfy hyperscaling, it follows that the highest
power in the expansion of the determinant
is determined by the dimensionality of space i.e. $p=d$. Consequently,
the upper critical dimensionality is
$d_c=p\sigma /(p-2)=d$. Thus, in the large-$N$ limit the mean-field treatment
is correct since any $d$ behaves as the critical dimensionality and the
Ginzburg
criterion is always respected in the sense of eq.(\ref{g9}).
Higher order $1/N$ corrections have two effects. They shift the critical
coupling and modify the exponents. While the first feature is expected
(critical couplings are nonuniversal), the second one might appear as a
surprise since different $N$ apparently produce different universality classes
which are not related to the symmetry group. This is another
feature of the effect of the long range forces.

Unlike local scalar theories, where non-gaussian scaling appears
only in a narrow region around the critical point, while mean-field
behavior prevails in a wider region away from it, in the Gross-Neveu model
non-gaussian behavior holds everywhere, close and away from the critical point,
and is exact in the $N\to \infty$ limit. This result of the $1/N$ expansion
has been corroborated by the lattice simulations of the model with 12
species \cite{HKK_93}. Beyond the $N=\infty$ limit
there is a small region in the vicinity of the transition point where the
$1/N$ deviations from the Gross-Neveu exponents
are observed. The width of that region
shrinks as some power of $1/N$ for large-$N$.
\vskip1truecm

\section
{ Simulations of the N=12 Model}

We have done simulations of the Gross-Neveu model in three and four dimensions
to test these ideas. Some of the three dimensional simulations have already
been reported, but here we will present the data and fits, discuss the results
and add new calculations. We begin with the four dimensional simulations
which are new. We chose to simulate the $Z_2$ four Fermi model because it
has been analyzed particularly thoroughly in the $1/N$ expansion and because
it is relatively easy to simulate. We simulate the model exactly with the
Hybrid Monte Carlo method for N=12. Using a large value for N suppresses
fluctuations and allows an accurate study within reasonable computer
resources. In addition, since the chiral symmetry is discrete the model
can be simulated directly in the chiral limit $m = 0.0$, so particularly
accurate determinations of the model's critical indices are possible. The
lattice action and further detail concerning the algorithm can be found
in ref.7.

\noindent
--{\bf Four Dimensions}

We simulated a $8 \times 16^3$ lattice as for the O(2) bosonic model, and
measured the vacuum expectation value of the auxiliary
$\sigma$ field and its susceptibility
$\chi$. The raw data is shown in Table 3.
The time step in the algorithm was $dt = 0.10$, the total
molecular dynamics 'time' expended at each coupling
$\beta$ ranged from 5,000 to
20,000. These numbers should be compared to typical finite temperature
lattice QCD simulations with light quarks where typically the statistics
is an order of magnitude less thorough. The error bars in the Table reflect,
as usual, the statistical uncertainties in the data accounting for
correlations in the usual way. As we discussed for the O(2) model above, we
cannot reliably simulate the model too close to the critical point
because of tunneling between $Z_2$ vacua on a finite system. So, although
we collected data at $\beta = .5875, .59$, and $.595$, we observed
tunneling in these data sets and could not use them for quantitative
purposes.

\noindent
--{\it Order Parameter}

We shall estimate the critical indices $\beta_{mag}$, $\gamma$ and $\delta$
in exactly the same fashion and using the same notation
that we did for the O(2) model on the
$8 \times 16^3$ lattice. In particular, we fit the $\sigma$ data over
the range of couplings .5700 - .5825 and found an excellent powerlaw
fit ( confidence level .995 ) which predicted the parameters $A = 1.6(2)$,
$\beta_c = .5910(1)$ and, most importantly,
the critical index $\beta_{mag} = .52(6)$. The fit
and the data are shown in Fig.5. Our measured value of $\beta_{mag}$ is
nicely consistent with mean field theory ( $\beta_{mag} = .50$ ), and it
is far from the three dimensional Ising value ( $\beta_{mag} = .325(2)$ )
predicted by the dimensional reduction scenerio. We feel that it is
interesting and significant that the powerlaw hypothesis fits the data so
well. This simplicity is expected from the analytic work we did above in
the sense that if the temperature regulates the theory's infra-red
fluctuations, then the theory's scaling laws should be particularly
accessible. If instead, the $8 \times 16^3$ lattice were too small and
insufficiently asymmetric to expose the true infra-red behavior of the
theory at finite temperature, then simple powerlaw scaling would be
quite a surprise. When we simulate the three dimensional Gross-Neveu
model at finite temperature we will study several lattice sizes and be
able to back up these impressions more quantitatively.

\noindent --{\it Susceptibility}

Next we fit our
susceptibility data with a simple powerlaw as was succesful for the
O(2) model and we found an excellent fit
(confidence level of .976) ,shown in Fig.6, but
the parameters were determined rather crudely ($A = 4.8(4.1)$, $C = -.1(1)$,
and $\gamma = .7(3)$ for $\beta_c = .5910$). This value of
the critical index $\gamma$ is
certainly consistent with the mean field value of $\gamma = 1.00$ and
disagrees with the three dimensional Ising value $\gamma = 1.241(2)$, but the
determination is crude and hardly persuasive. As usual, it proves harder
to measure susceptibilities than order parameters in simulations and the
susceptibilities are subject to larger finite size effects.

\noindent --{\it $\delta$ Exponent}

Now we turn to our determination of the index $\delta$. We measured the
order parameter $\sigma$ at the fermion masses
.002, .004, .006, .008, and .010
at the critical coupling $\beta_c =.5910$. The data is collected in Table 4.
Powerlaw
fits to the m-dependence, using the same notation as for the O(2) model of
Sec.3,
produced a very good fit (confidence level of .80 ), and predicted the
parameters $A = .969(9)$, and the critical index $\delta = 3.23(3)$. The value
of $\delta$
obtained here is close to the mean field value of 3.00, and we interpret
the discrepancy between the two numbers as a finite size effect. It is
well-known that estimates of $\delta$ coming from finite lattice simulations
produce numbers which are systematically higher than the actual value and
the correct value is obtained only in the thermodynamic limit. Nonetheless,
the important point is that the $\delta$ exponent obtained here is
far from that of the O(2) model simulated also on a $8 \times 16^3$
lattice ( recall from the previous section that $\delta = 4.7(2)$ in
that case ). The data and the fits for the $Z_2$ Gross-Neveu model are
shown in Fig.7 and 8. In Fig.9 we combine Fig.4 and 8, showing $ln(\sigma)$
vs. $ln(m)$ for the two cases. In both cases the powerlaw ansatz
holds well but the predicted values of $\delta$ are very different - the
O(2) model satisfies dimensional reduction while the four Fermi model
does not. Naturally it would be good to repeat these sorts of simulations
on larger lattices and do a thorough finite size scaling analysis of
both models.

We have argued elsewhere that a particularly visual way of looking  for
approximate mean field scaling is to plot the theory's 'transverse
susceptibility', $m/\sigma$, against the square of the order parameter
at criticality.
In mean field theory such a plot would be a straight line passing through
the origin. In Fig.10 we show such a plot for the data in Table 3. The
plot has slight curvature because our measured critical indices differ
slightly from mean field theory. To the accuracy of our data
and its fit, the dashed
curve passes through the origin.
\noindent

--{\bf Three Dimensions}

The conceptual issues discussed in this paper can be challenged more
persuasively by considering the three dimensional $Z_2$ Gross-Neveu
model at finite temperature and examining whether its critical
behavior is that of the two dimensional Ising model, as predicted
by the dimensional reduction scenerio, or mean field theory, as
predicted by the large-$N$ solution of the model. In three dimensions
we can simulate larger lattices with greater statistics so the model's
critical indices can be determined with greater accuracy and confidence.
Finite size scaling studies are more feasible and the width of the
critical region at nonzero temperature can be examined numerically. And
lastly, the critical indices and universal amplitude ratios
of the two dimensional Ising model are
{\it far} from mean field theory, so the two scenerios can be distinguished
numerically with greater confidence than in four dimensions.

Several studies have been made of the $Z_2$ Gross-Neveu model in three
dimensions
and the predictions of the $1/N$ expansion have been reproduced by
the same simulation methods as used here. The zero temperature, large-$N$
critical indices of the three dimensional Gross-Neveu model are recorded in
Table 5. As emphasized in the text and is well-known to workers in the field,
the fixed point of the three dimensional Gross-Neveu model is {\it different}
from that of the $Z_2$ $\phi^4$ model. The physical origin of the difference
lies in the fact that the fermionic Gross-Neveu model cannot be expressed
as a {\it local} bosonic model with the same symmetries. Therefore, it can
and does determine a 'new' universality class, which at large-$N$ has the
critical indices listed in the table. The relation between the two fixed
points has been studied in detail within the context of Yukawa models.
We also list in the table the mean field exponents expected to describe
the finite temperature transitions in Gross-Neveu models in any dimension, as
well as the two dimensional Ising model indices which should apply to
the finite temperature transition of a bosonic model in three dimensions
which satisfies dimensional reduction. As we discussed in a recent letter on
this subject \cite{KK_95}, the mean field and the two dimensional Ising
exponents are
distinctly different. In particular the two dimensional Ising model has
$\delta = 15$ which is five times larger than the mean field prediction!

Notice from the table that the critical indices $\beta_{mag}$, $\delta$, and
$\gamma$ do not change monotonically as we pass from the $d = 3$ Gross-Neveu
model to the $T \neq 0$ Gross-Neveu model to the two dimensional Ising model :
while $\beta_{mag}$ decreases from $1$ to $1/2$ to $1/8$ and $\delta$ increases
from $2$ to $3$ to $15$, the index $\gamma$ varies from $1$ to $1$ to $7/4$.
So, if we confirm the mean field exponents in our finite temperature
simulations, it will not be easy to interpret the results as evidence that
'our lattices are too small and not sufficiently aymmetric to measure
the true finite temperature indices'. In addition, we will do simulations
on several lattice sizes to check that we are in the scaling region of the
finite temperature transition of interest. The lattices themselves will
have many more sites per dimension than typical four dimensional simulations.
\vskip5truemm
\noindent
{\bf 6.1 $6\times 30^2$ Lattice}

We begin with our simulations on $6 \times 30^2$ lattices. This lattice size
was chosen because its asymmetry ratio is large ( $30/6$ ) and because our
past studies of the zero temperature model indicated that the lattice is just
barely
large enough to obtain critical indices and powerlaw scaling laws with
good accuracy \cite{HKK_93}. Lattices which are smaller than this in the
'temperature'
direction would produce results badly distorted by discreteness. We present
the data in Table 6. The notation is the same as previous tables. The
statistics
for each data point are far better than the four dimensional simulations -
more than 100,000 sweeps were done at each parameter setting in each of
our three dimensional simulations, unless stated otherwise. The notation
for our fits will follow the same conventions as we used in the four
dimensional study presented above. First, consider the order parameter
and its critical index $\beta_{mag}$. A simple powerlaw fit
over the range of couplings $.70$ through $.76$ works well and
predicts the parameters $A = 1.33(2)$ and $\beta_{mag} = .52(2)$ in fine
agreement with mean field theory. The data and the fit are shown in
Fig.11. (The figure also includes the data and fit from a simulation
on a much larger lattice $12 \times 36^2$ which will be discussed
below.) The critical coupling on the $6 \times 30^2$ lattice
is determined to be $.792(1)$. Next we considered the susceptibility
data and fit $1/\chi$ as discussed earlier in Sec.3. The range of couplings
$.68$
through $.75$ produced an adequate powerlaw fit with the parameters
$A = 12.9(1.3)$, $C = -.095(65)$, and the critical index $\gamma = .99(7)$.
The data and its fit are shown in Fig.12 along with the $12 \times 36^2$
data. As usual, susceptibility data and fits are considerably less accurate
than order parameter studies. Nonetheless, the results are in fine
agreement with mean field theory and contrast sharply with the two
dimensional Ising value of $\gamma = 7/4$. Finally we consider the model at
criticality and estimate the index $\delta$. The data is given in Table 7
and powerlaw fits over the mass range of $.00375$ - $.015$ produce
the parameters $A = .89(1)$ and $\delta = 3.40(4)$. The data and its fit
are plotted in Fig.13 along with the $12 \times 36^2$ results. As
expected the measured $\delta$ is slightly higher than the 'expected' mean
field result due to finite size effects. Nonetheless, since the two
dimensional Ising value is $\delta = 15$, the numerical distinction
between the two scenerios is brilliantly clear. It is significant that
simple powerlaw fits were adequate in this application. This result
gives more credence to the fact that the simulation is detecting the
true finite temperature scaling law of the model and is not seriously
distorted by a possible sluggish cross-over between the three dimensional
Gross-Neveu model and the two dimensional Ising model.
\vskip5truemm

\noindent
{\bf 6.2 $12\times 36^2$ and $12\times 72^2$ Lattices}

Now we turn to the $12 \times 36^2$ simulations and extensive results on
a relatively huge $12 \times 72^2$ lattice. The $12 \times 36^2$ data is given
in Tables
8 and 9, in the same format as above. We simulated these larger lattices to
gather further evidence that we are measuring properties of the finite
temperature behavior of the model. The larger 'temperature' extent of the
lattice should remove some of the distortion in the $6 \times 30^2$
lattice due to discreteness. The asymmetry ratio of $36/12$ should be
sufficient to resolve the zero temperature, bulk trasition reviewed
in Table 5 from the true finite temperature transition. If the $6 \times
30^2$ results were just indicative of a sluggish cross-over rather than
a true finite temperature transition, the the $12 \times 36^2$ results
should be distinctly different. We shall find no evidence for this view.
Instead, the $12 \times 36^2$ results will be in good agreement with
the $6 \times 30^2$ results and mean field theory.

\noindent
--{\it Order Parameter and $\beta_{mag}$}

In Fig.11 we show the data and its fit for the order parameter as the
coupling varies. The fit is done over the range of couplings $.80$ through
$.85$. Its confidence level is .78 and its parameters are $A = 1.13(4)$,
and $\beta_{mag} = .58(3)$ for $\beta_c = .881$. The critical index
$\beta_{mag}$ is slightly above the $6 \times 30^2$ and the mean field
prediction of $1/2$. This small discrepancy is probably due to the smaller
asymmetry ratio used here. The result is far from the two dimensional
Ising value of $1/8 = .125$. Next we fit the susceptibility data
over the coupling range $.82$ through $.85$ and determine the parameters
$A = 8.5(2.1)$ and $\gamma = .9(1)$ with a confidence level of .17. The
critical index $\gamma$ is in good agreement with the $6 \times 30^2$
simulation and with mean field theory. The data and the fit have already
been plotted in Fig.12. Even though the uncertainties are
larger than we would have liked, the results are clearly distinct from the
Ising value of $7/4$. Finally, we simulated the model at the critical
coupling and determined the index $\delta$. A powerlaw fit for the
range of bare fermion masses $.0009$ through $.0100$ produced the parameters
$A = .91(3)$ and $\delta = 2.90(6)$ with confidence level .13. The data and
the fit have already been shown in Fig.13. If we fit the data over the
mass range $.0009$ through $.0025$, a slightly better fit results ( confidence
level .22 ) with the parameters $A = .94(9)$ and $\delta = 2.8(2)$. In either
case the results are in good agreement with mean field theory and support
the analytic large-$N$ results.

Finally, we checked the $12 \times 36^2$ results with
simulations on a $12 \times 72^2$ lattice. The data are given in Table 10. The
results
for the order parameter $\sigma$
are within a standard deviation of the $12 \times 36^2$ results and they
support the view that the lattices studied here are large enough and
asymmetric enough to detect the true finite temperature properties of
the theory. Note that $\sigma =.1020(10)$ at $\beta=.86$. Since $\sigma$
is also the dynamical fermion mass at large-$N$, and since the lowest
scalar mass at large-$N$ is $m_{\sigma} = 2\sigma$, the largest
correlation length in this data set is approximately 5 lattice spacings.
Vacuum tunneling made it impossible for us to simulate the model
closer to the critical point than this. However, in the {\it symmetric}
phase we certainly simulated larger correlation lengths. As discussed
briefly in the Introduction, in the 2-d Ising model as well as in mean
field theory, the correlation function and the susceptibility in the
symmetric phase are closely related to these same quantities in the
broken phase. In the 2-d Ising model, the correlation function amplitude
ratio is universal and is $2$, and the susceptibility amplitude ratio
is also universal and is 37.69. In mean field theory these universal
amplitude ratios are $2^{1/2}$ and $2$, respectively. We will see below
the $12 \times 72^2$ data predicts that the critical coupling is very
near $.87$, so if our data is described by the 2-d Ising model then the
correlation length at $\beta = .88$ would be approximately $10$ while if
mean field theory applies it would be approximately $7$ there. In either case
these are substantial correlation lengths, so it would not be foolish to
expect the data to exhibit the true critical behavior of the
finite temperature transition.

Since the $12 \times 72^2$ data for the order parameter agrees with our
$ 12 \times 36^2$ data, we should extract the same index $\beta_{mag}$ as
before. The $12 \times 72^2$ data is, however, slightly more accurate
and closer to the critical point. Powerlaw fits to the $\sigma$ vs.
$\beta$ data predict $\beta_{mag} = .53(2)$, $A = 1.00(3)$ and a critical
coupling $\beta_c = .875$ with a confidence level of .53. The agreement
with mean field theory is almost perfect and the 2-d Ising model
prediction of $\beta_{mag} = 1/8$ is easily ruled out.

\noindent
--{\it Susceptibility and $\gamma$}

Measurements of the theory's susceptibility suffer from larger statistical
errors although typically 100,000 sweeps of the algorithm were run at
each parameter setting near the critical coupling. Powerlaw fits to the
susceptibility in the broken phase predicted a critical index of
$\gamma = .72(14)$, $A = 6.0(1.7)$ and a critical coupling $\beta_c =
.865$ with a confidence level of .56. The result for $\gamma$ is slightly
more than one standard deviation below the prediction $\gamma = 1.00$ of mean
field theory and is far from the 2-d Ising model's $\gamma = 1.75$. The
susceptibility data in the symmetric phase is slightly better because of
the absence of tunneling and the larger correlation lengths. Here the fits
predict $\gamma = 1.17(15)$, $A = 7.38(3.23)$ and a critical coupling
of $\beta_c = .870$ with a confidence level of .92.

\noindent
--{\it Universal Amplitude Ratios}

Perhaps our most persuasive evidence for mean field behavior follows from
the universal susceptibility ratio. Mean field theory and the 2-d Ising model
differ by a factor approaching 20 for this ratio so even crude data within
the theory's scaling region will distinguish the two alternatives. In Fig.14
we show the inverse susceptibility and the mean field fit assuming
$\gamma = 1.0$. The fit contains only two free parameters -- the critical
coupling (which must be near .87 to agree with other fits) and the
amplitude (in the broken phase). The resulting figure looks fairly compelling
and should be compared to Fig.15 where the same procedure has been
tested assuming the applicability of the 2-d Ising model. That figure
shows that the amplitude ratio of 37.69 characteristic of the 2-d Ising
model utteringly fails.
\vskip5truemm

\section
{Simulations of the N=4 Model}

Now we turn to extensive simulations of the model with one-third the
number of fermion species. Our motivation for setting $N$ to four is to
investigate
whether mean field predictions only apply to the large-$N$ limit
of the model. The following secenerio has been suggested in Ref.17 based
on experience with large-$N$ scalar models: Since $1/N$ is a dimensionless
quantity, it might determine the true width of the scaling region of the model.
At infinite $N$ mean field behavior could apply in a region of width of order
unity around $\beta_c$. However, at finite $N$ a region of width
O(1/N) might emerge around the critical point inside of which
dimensional reduction applies and the true
scaling behavior is that of the 2-d Ising model. For a finite extent outside
the 'true' scaling region, mean field behavior would apply as predicted
by the large-$N$ gap equation. Outside that region the correlation length
would be sufficiently small that the zero temperature critical point would
control the theory's scaling.
As discussed above, we have no theoretical reasons to suspect that
this scenerio applies to fermionic models and the physical picture
developed in this article does not hint at any subtle $N$ dependence.
Nonetheless,
in the absence of a rigorous solution to the model at finite $N$, we must
address
this question computationally. Luckily, the computer algorithm
works well for any $N$. Of
course, at small $N$ we expect more serious fluctuations and vacuum tunneling
so the computer simulations will have their usual limitations. We shall
try to master these practical problems with long simulations on relatively
large lattices, but some additional, more sophisticated studies would also be
welcome.

Extensive simulations of the $N=4$ model were run on $6 \times 36^2$ and
$12 \times 72^2$ lattices. In Table 11 we collect
the $6 \times 36^2$ data for $\sigma$ and
its susceptibility for couplings $\beta$ ranging from .60 to .80. To achieve
useful accuracy 250,000 sweeps of the algorithm were run at $\beta$ values
near the transition (from .64 through .72) while 100,000 sweeps were
accumulated elsewhere. We could not collect useful data at $\beta$ values
.66, .67 or .68 because of vacuum tunneling ( Our $12 \times 72^2$
simulations will improve on this limitation. ). The error bars in the table
account for correlations in the raw data sets. Even with longer runs
near the critical point, the error bars increase near criticality in Table 11.

\noindent
--{ $<\bar\psi\psi>$}

First consider the critical index $\beta_{mag}$. In Fig.16 we plot $\sigma^2$
vs. $\beta$. The curve is well-approximated by a straight line, especially
near the critical point, as expected if mean field theory applies. A
powerlaw fit predicts $\beta_{mag} = .4(1)$, $A = 1.31(9)$ and critical
coupling $\beta_c = .67$ with a confidence level of .39.

\noindent
--{\it Susceptibility}

Next consider the susceptibility results. In Fig.17 we plot the inverse
susceptibility vs. coupling. We are particularly interested in determining
the critical index $\gamma$ here and in seeing whether the data is compatible
with the universal amplitude ratio of $2$ expected of mean field theory. The
dashed lines in the figure show the fit which incorporates
$\gamma = 1.0$. The fit used the data in the broken phase close
to the critical point, at $\beta = .64$ and $.65$, to determine the
susceptibility amplitude and critical coupling, and it then predicted the
susceptibility in the symmetric phase.
We see from the figure that the fit reproduces the
susceptibility quite well.
Again, the $12 \times 72^2$ simulation will yield more useful data
near the critical point, so the fit will be even more
compelling.

The extensive susceptibility data in the symmetric phase allowed us to test
the mean field prediction of $\gamma = 1.0$ directly. Powerlaw fits to the data
gave $\gamma = 1.2(2)$, $A = 1.6(8)$, and $\beta=.670$ with a confidence
level of .57. The 2-d Ising model value of 1.75 for $\gamma$ is not
compatible with the data. To emphasize this point we show in Fig.18 the
2-d Ising model fit to the susceptibility data incorporating $\gamma_{Ising}
= 1.75$ and the universal amplitude ratio 37.69. As before we fit to the
data near the critical point in the broken phase
and then 'predicted' the
data in the broken phase. We see that the 'prediction', the dashed line
in the figure, does not resemble the data.
The only limitation of this fit is the fact
that we do not have much data in the scaling region on the broken coupling side
of the transition. On this size lattice we are simply unable to avoid
considerable vacuum tunneling.

Finally, consider the $12 \times 72^2$ data for the $N = 4$ model. The data is
shown in Table 12. As usual, the error bars reflect correlations in the
data set. The runs were as long as 250,000 sweeps near the critical point. Note
that the order parameter is .164(5) at $\beta=.745$. We estimate, therefore,
that the scalar correlation length is roughly 3.05 at this coupling. We
could not simulate the model in the broken phase closer to the critical
point because of vacuum tunneling, so we took extensive measurements in
the symmetric phase to produce data with larger correlation lengths. ( We shall
see that within mean field fits, the correlation length at $\beta = .76$
is roughly 7.7. ). There were signs of vacuum tunneling in the $\beta = .745$
data sets, and this accounts for the relatively large error bars for
the order parameter and its susceptibility there. Luckily, the tunneling
events were rare enough in the data that they could be isolated and
discarded.

\noindent
--{$<\bar\psi\psi>$}

First consider the dependence of the order parameter on the coupling. We show
$\sigma^2$ vs. $\beta$ in Fig.19, and the linearity of the fit clearly
supports mean field theory. Powerlaw fits to the data for $\beta$
ranging from $.70$ through $.745$ predict $\beta_{mag} = .51(4)$, in almost
perfect aggreement with the mean field scenerio. The remaining parameters
of the fit are $A = 1.22(6)$, $\beta_c = .7635$, and the confidence level
of the fit is .67.

\noindent
--{\it Susceptibility}

Next consider the susceptibility data. The error bars on the susceptibility
data are larger than we would like, so fits to determine the index
$\gamma$ are not very informative. Powerlaw fits to the data in the broken
phase give $\gamma = .8(4)$ and fits in the symmetric phase give
$\gamma = 1.4(2)$. Comparisons of the amplitude ratios to the 2-d Ising and
mean field prediction are, however, much cleaner. In Fig.20 we show
the inverse of the susceptibility vs. coupling. The dashed line
gives the mean field prediction for the entire curve based on just two
parameters, $\beta_c$ and the amplitude in the broken phase. The curve is
in general agreement with the data for a very reasonable $\beta_c = .764$
which is nicely compatible with the order parameter data. In Fig.21
we show the results of the same exercise using the fitting form of the
2-d Ising model. We see that if we fit the susceptibility data in the
broken phase near $\beta_c$, then we fail qualitatively to describe the
data in the symmetric phase. One might try to salvage the 2-d Ising
scenerio by moving the estimate of $\beta_c$ to a smaller value in order
to accommodate the susceptibility data in the symmetric phase at the
expense of a poorer 'fit' in the broken phase. In Fig.22 we show the
case with $\beta_c = .758$. The description of the data remains unsuccessful.

\noindent
--{\it $\delta$ Exponent}

Finally, we turn to a determination of the critical index $\delta$. This is
a useful universal quantity to calculate because is differs by a factor
of five in the two scenerios. The data for the order parameter at
criticality ( we estimated $\beta_c = .760$ for these runs ) vs. the bare
fermion mass is shown in Table 13. Note that our lowest $\sigma$ value
in the Table is .0856 which corresponds to a correlation length of
5.8. A plot of $\ln(1/\sigma)$ vs. $\ln(1/m)$ should have a slope of
$1/\delta$. The data is plotted this way in Fig. 23. The darker dashed
line is the mean field prediction of $\delta = 3.0$ and the lighter
line is the 2-d Ising prediction of $\delta = 15$. The data with the
largest correlation lengths are compatible with $\delta = 3.0$ while
none of the data points resemble the 2-d Ising scenerio.

Of course, one can always imagine doing more simulations on
larger lattices with greater statistics and resolution, complete with
a thorough finite size scaling analysis. Perhaps the results presented here
will inspire such activity. Certainly our simulations strongly favor
the physical scenerio presented above where fermions are essential at
the finite temperature transition and dimensional reduction does not
occur.

\section
{ Conclusions}

We have studied the universality class of the chiral restoration transition
in the $Z_2$ Gross-Neveu model at finite temperature using large-$N$ and
lattice simulation. Our findings strongly support mean field scaling
instead of that of a dimensionally reduced Ising model. As we mentioned in
the introduction, we found our initial findings, reported in \cite{KK_95},
perplexing
and, to some extent, circumstantial, and so we continued and refined our study
in order to close possible loopholes in
our past results. In particular, returning to
the three points from the Introduction, we addressed the question of the
narrowness of the scaling region and the possible failure of the
$1/N$ expansion in this context. The leading order calculations do not
hint at any departure from mean field behavior. However, it is possible
that non-Gaussian scaling is resurrected beyond the leading order. If that
were the case, there should be a scaling region whose width scales
as some power of $1/N$ in which the deviations from mean field scaling are
visible.
Due to numerical limitations, it is conceivable that the $N=12$ simulations are
not sufficiently good to enter that region. For that purpose we simulated the
$N=4$ theory. We observed no change in our conclusions. Our lattices were
sufficiently large to see any such departures if they were there. After all,
in past simulations the
correct physics was recovered when the 2d Gross-Neveu model was studied
on lattices of similar extent \cite{KKW_87}.

Our task of excluding the dimensional reduction scenario for the 3d Gross-Neveu
model was relatively
easy. This was so because the exponents $\delta$ and $\beta_{mag}$ differ
by factors of
5 and 4, respectively, in the mean field and 2d
Ising models. Furthermore, the susceptibility amplitude ratios
differ by almost a factor of 20 in these two cases and there is virtually no
chance they can be confused with each other even on modest lattice sizes
studied with modest statistics.

In Table 5 we record the critical indices of the 3d large-$N$ Gross-Neveu
model,
mean field indices and those of the 2d Ising model.
If the mean field scaling we observed is
just an effect of a sluggish crossover from $T=0$ to 2d Ising scaling,
then all the exponents should experience a uniform approach to the Ising
values. From the table it is clear that there is no such uniformity.
The 'crossover' interpretation of our simulation results is hard to reconcile
with this observation.

To eliminate any doubts about observing dimensional reduction
in lattice simulations, we simulated the scalar $O(2)$ model where the
standard scenario seems inevitable. We undertook this in order to
get some ideas about the lattice sizes and correlation lengths needed for
the dimensional reduction scenerio to manifest itself. We found
scaling consistent with the dimensional reduction
on lattices of the modest size, considerably smaller than we
used for the Gross-Neveu model.

Therefore, we conclude that the finite temperature
scaling of the finite temperature Gross-Neveu is not that of a dimensionally
reduced model.

As far as $QCD$ is concerned, the moral of the Gross-Neveu exercise is
twofold. First, from the physics point of view, four-fermi models are closer
in spirit to $QCD$ than the $\sigma$-models since they both realize mesons as
quark composites. Therefore, they share this apparently crucial aspect. It is
conceivable, but by no means inevitable that the $QCD$ transition is in the
same universality class as the Nambu-Jona-Lasinio model \cite{NJL_61}.
At least, there is no
apriori argument against it. From the numerical point of view, the Gross-Neveu
exercise is illustrative of the degree of difficulty involved in addressing
questions like the universality class of the theory. This is especially
relevant in light of the fact that the question of what symmetry gets
restored \cite{Shu_94}
at finite temperature lattice $QCD$ requires computer power
capable of distinguishing the exponents of the $O(2)$ and $O(4)$ models,
which are
within a few percent of each other \cite{Kar_94}. We have had
some success in studying this same question in a different, simpler context:
our task
was to distinguish mean field scaling laws from those of the 2d Ising
model where differences of a factor of 20 occur.
Nevertheless, this proved to be a nontrivial task and there might
still be some scepticism remaining about our findings. In this light
the question about the universality class of $QCD$ remains one that will
be open for quite a while. Small lattice simulations of the finite temperature
transition in QCD cannot distinguish between mean field, $O(2)$, or $O(4)$
behavior at this time \cite{Kar_94}.
Perhaps some of the observations and methods of
analysis reported in this paper will
help decide this important question. Certainly,
before one studies the dynamics of the
chiral transition and heavy ion phenomenology,
one needs to determine the universality class of the static transition.

We wish to acknowledge discussions with Misha Stephanov.
This work is supported by NSF-PHY 92-00148
and used the computing facilities of PSC and NERSC.
\vfill\eject

\newpage\noindent
\begin{thetable}

\begin{table}
\caption[]{Four-dimensional scalar $O(2)$ model
on a $8\times 16^3$ lattice}
\begin{tabular}{ccc}
$\beta$ & $\sigma$ & $\chi$\\
0.19   & 0.646(1) & 0.695(5) \\
0.185  & 0.622(1) & 0.835(3) \\
0.18   & 0.595(1) & 1.023(4) \\
0.175  & 0.561(1) & 1.302(3) \\
0.17   & 0.421(1) & 1.761(7) \\
0.165  & 0.469(1) & 2.569(13)\\
0.1625 & 0.437(1) & 3.221(7) \\
0.16   & 0.400(1) & 4.313(44)\\
0.1575 & 0.353(1) & 6.334(31)\\
0.155  & 0.292(1) & 10.884(52)\\
0.1525 & 0.199(1) & 31.76(62) \\
\end{tabular}
\end{table}

\begin{table}
\caption[]{Order parameter at criticality for the four-dimensional scalar
$O(2)$
model on a $8\times 16^3$ lattice}
\begin{tabular}{ccc}
$m$ & $\sigma$ \\
0.002   & 0.219(2) \\
0.003   & 0.235(1) \\
0.004   & 0.249(1) \\
0.005   & 0.260(1) \\
0.006   & 0.271(1) \\
0.007   & 0.279(1) \\
0.008   & 0.287(1) \\
0.009   & 0.295(1) \\
0.01    & 0.302(1) \\
0.015   & 0.332(1) \\
0.02    & 0.354(1) \\
0.03    & 0.390(1) \\
0.04    & 0.417(1) \\
0.05    & 0.440(1) \\
\end{tabular}
\end{table}

\begin{table}
\caption[]{Data compilation for the four-dimensional
$Z_2$ four-Fermi model on a $8\times 16^3$ lattice}
\begin{tabular}{ccc}
$\beta$ & $\sigma$ & $\chi$  \\
0.56 & 0.2660(3) & 2.80(15) \\
0.565& 0.2420(4) & 3.28(15) \\
0.57 & 0.2159(3) & 4.18(12) \\
0.5725& 0.2020(3)& 4.73(9)  \\
0.575& 0.1870(4) & 5.48(15) \\
0.5775& 0.1711(5)& 6.71(21) \\
0.58 & 0.1534(7) & 8.80(39) \\
0.5825& 0.1336(9)& 12.48(79)\\
\end{tabular}
\end{table}

\begin{table}
\caption[]{Order parameter at criticality 2 for the four-dimensional
$Z_2$ four-Fermi model on a $8\times 16^3$ lattice}
\begin{tabular}{ccc}
$m$ & $\sigma$ \\
0.002   & 0.1415(5) \\
0.004   & 0.1757(4) \\
0.006   & 0.1994(3) \\
0.008   & 0.2177(2) \\
0.010   & 0.2332(2) \\
\end{tabular}
\end{table}

\begin{table}
\caption[]{Critical indices for the models of interest}
\begin{tabular}{|c|c|c|c|} \hline
{} &
\multicolumn{2}{c|}{3d  Gross-Neveu} &
2d Ising \\ \cline{2-3}
{} & $T=0$ & $T\not= 0$ & {} \\ \hline
$\beta$ & 1 & 1/2 & 1/8 \\ \hline
$\delta$ & 2 & 3 & 15 \\ \hline
$\gamma$ & 1 & 1 & 7/4 \\ \hline
$\nu $ & 1 & 1/2 & 1 \\ \hline
$\eta $ & 1 & 0 & 1/4 \\ \hline
\end{tabular}
\end{table}

\begin{table}
\caption[]{Data compilation for the order parameter, $\sigma$ and its
susceptibility, $\chi$, versus the coupling, $\beta$ in the $Z_2$ four-Fermi
model on a $6\times 30^2$ lattice}
\begin{tabular}{ccc}
$\beta$ & $\sigma$ & $\chi$  \\
0.64& 0.5116(2) & 0.545(3) \\
0.66& 0.4701(2) & 0.624(6) \\
0.68& 0.4276(2) & 0.737(3) \\
0.69& 0.4061(3) & 0.799(5) \\
0.70& 0.3836(2) & 0.897(3) \\
0.71& 0.3608(4) & 1.023(5) \\
0.72& 0.3368(4) & 1.14(2)  \\
0.73& 0.3119(4) & 1.405(5) \\
0.74& 0.2847(4) & 1.68(2)  \\
0.75& 0.2557(8) & 2.26(5)  \\
0.76& 0.2199(5) & 3.60(10) \\
0.77& 0.1734(13)& 8.47(1.00)\\
\end{tabular}
\end{table}

\begin{table}
\caption[]{Data compilation for $\sigma$ vs. $m$, the bare fermion mass,
at the critical point on a $6\times 30^2$ lattice for the $Z_2$
four-Fermi model}
\begin{tabular}{ccc}
$m$ & $\sigma$ \\
0.00125 & 0.1080(13) \\
0.002   & 0.1358(11) \\
0.0025  & 0.1489(7)  \\
0.00375 & 0.1719(9)  \\
0.005   & 0.1885(5)  \\
0.0075  & 0.2131(5)  \\
0.01    & 0.2317(3)  \\
0.015   & 0.2595(4)  \\
0.02    & 0.2811(1)  \\
0.03    & 0.3136(1)  \\
0.04    & 0.3373(1)  \\
\end{tabular}
\end{table}

\begin{table}
\caption[]{Same as Table 6 on a $12 \times 36^2$ lattice}
\begin{tabular}{ccc}
$\beta$ & $\sigma$ & $\chi$  \\
0.70& 0.4299(8) & 0.729(7)  \\
0.72& 0.3943(2) & 0.736(9)  \\
0.74& 0.3591(2) & 0.791(9)  \\
0.76& 0.3233(4) & 0.934(20) \\
0.78& 0.2871(5) & 1.145(20) \\
0.79& 0.2693(4) & 1.220(20) \\
0.80& 0.2499(4) & 1.279(15) \\
0.81& 0.2304(3) & 1.535(15) \\
0.82& 0.2108(8) & 1.784(10) \\
0.83& 0.1884(3) & 2.160(20) \\
0.84& 0.1645(5) & 2.920(20) \\
0.85& 0.1384(5) & 4.29(10)  \\
\end{tabular}
\end{table}

\begin{table}
\caption[]{Same as Table 7 on a $12\times 36^2$ lattice}
\begin{tabular}{ccc}
$m$ & $\sigma$ \\
0.0009  & 0.0798(11) \\
0.00125 & 0.0951(58) \\
0.0018  & 0.1051(19) \\
0.0025  & 0.1148(5)  \\
0.005   & 0.1475(9)  \\
0.010   & 0.1834(15) \\
\end{tabular}
\end{table}

\begin{table}
\caption[]{ $12 \times 72^2$ lattice, $N=12$}
\begin{tabular}{cccc}
$\beta$ & $\sigma$ & $\chi$ & sweeps ($\times 1000$) \\
0.80& 0.2499(1) & 1.29(2) & 40 \\
0.81& 0.2304(3) & 1.54(5) & 40 \\
0.82& 0.2101(4) & 1.72(5) & 20 \\
0.83& 0.1884(3) & 2.16(9) & 20 \\
0.84& 0.1645(4) & 2.82(12)& 50 \\
0.85& 0.1370(5) & 4.54(14)& 50 \\
0.86& 0.1020(10)& 12.0(1.0)&50  \\
0.88&           & 24.40(84)&80  \\
0.89&           & 11.91(5) &70  \\
0.90&           & 7.83(47) &70  \\
0.91&           & 5.51(28) &80  \\
0.92&           & 4.35(27) &80  \\
0.93&           & 3.41(13) &50  \\
\end{tabular}
\end{table}

\begin{table}
\caption[]{ $6 \times 36^2$ lattice, $N=4$}
\begin{tabular}{cccc}
$\beta$ & $\sigma$ & $\chi$ & sweeps ($\times 1000$) \\
0.60& 0.4869(8) & 3.51(12) & 90 \\
0.61& 0.4575(6) & 4.00(16) & 90 \\
0.62& 0.4271(9) & 4.86(16) & 90 \\
0.63& 0.3933(9) & 6.34(21) & 90 \\
0.64& 0.3514(17) & 11.8(1.8)& 250 \\
0.65& 0.3029(21) & 17.6(1.3)& 250 \\
0.69&           & 59.7(2.9) &250 \\
0.70&           & 39.5(1.6) &250  \\
0.71&           & 27.5(1.3) &250 \\
0.72&           & 20.62(79) &250  \\
0.73&           & 17.70(94) &90  \\
0.74&           & 14.77(77) &90  \\
0.75&           & 11.75(54) &90  \\
0.76&           & 9.87(52) &90  \\
0.77&           & 8.60(28) &90  \\
0.78&           & 7.62(30) &90  \\
0.79&           & 6.90(27) &80  \\
0.80&           & 6.13(22) &80  \\
\end{tabular}
\end{table}

\begin{table}
\caption[]{ $12 \times 72^2$ lattice, $N=4$}
\begin{tabular}{cccc}
$\beta$ & $\sigma$ & $\chi$ & sweeps ($\times 1000$) \\
0.70& 0.3061(2) & 4.45(7) & 60 \\
0.71& 0.2813(7) & 6.48(15) & 120 \\
0.72& 0.2534(3) & 7.32(39) & 180 \\
0.73& 0.2238(11) & 9.96(49) & 120 \\
0.74& 0.1866(15) & 17.3(1.2)& 140 \\
0.745& 0.164(5) & 27.9(5.3)& 100 \\
0.77&           & 70.5(9.1) & 140  \\
0.78&           & 42.2(1.7) & 140  \\
0.79&           & 25.1(2.3) & 100 \\
0.80&           & 20.2(1.4) & 120 \\
0.81&           & 18.1(1.3) & 220 \\
0.82&           & 13.01(86) & 120 \\
0.83&           & 10.29(29) & 100 \\
0.84&           & 9.42(57) & 100 \\
0.85&           & 8.02(75) & 60 \\
\end{tabular}
\end{table}

\begin{table}
\caption[]{ $12 \times 72^2$ lattice, $N=4$ at Criticality}
\begin{tabular}{cccc}
$m$ & $\sigma$ & $\chi$ & sweeps ($\times 1000$) \\
0.00025& 0.0856(78) & 99.2(9.0) & 240 \\
0.0005& 0.1077(24) & 43.5(9.9) & 200 \\
0.001& 0.1363(13) & 21.5(1.3) & 320 \\
0.002& 0.1637(17) & 13.9(7) & 200 \\
0.003& 0.1820(8) & 8.16(43) & 200 \\
0.004& 0.1946(9) & 7.63(30) & 200 \\
0.005& 0.2058(2) & 6.48(43) & 80 \\
\end{tabular}
\end{table}

\end{thetable}
\newpage\noindent
\section*{Figure Captions} \begin{enumerate}

\item Order parameter vs. coupling for $O(2)$ model on an $8 \times 16^3$
lattice. The
dashed line is the fit discussed in the text.

\item Same as Fig. 1 except for the inverse susceptibility

\item Order parameter vs. mass for $O(2)$ model on an $8 \times 16^3$ lattice
at the critical coupling. The dashed line is the fit discussed in the text.

\item log-log plot of the data in Fig. 3.

\item Same as Fig.1 except for the Nambu Jona-Lasinio model.

\item Same as Fig. 5 except for the inverse susceptibility.

\item Same as Fig.3 except for NJL model.

\item log-log plot of the data in Fig. 7

\item Figs. 4 and 8 combined to illustrate the different slopes (critical
indices
$\delta$).

\item Tranverse susceptibility ($m/\sigma$) vs $\sigma^2$ for the NJL model at
the critical coupling on a $8 \times 16^3$ lattice. A straight line passing
through the origin would agree exactly with mean field theory.

\item Order parameter squared vs. coupling for the NJL model on $6 \times 30^2$
(lower curve) and $12 \times 36^2$ lattices.

\item Same as Fig. 11 except for the inverse susceptibility.

\item Log-log plot of the order parameter vs. mass at the critical point for
both
the $6 \times 30^2$ lattice (lower set of data points and their fit discussed
in the text) and the $12 \times 36^2$ lattice. The dashed-dotted curve is the
2-d
Ising model prediction.

\item Inverse susceptibility $\chi$ vs. coupling for the NJL model on a $12
\times 72^2$
lattice. The dashed curve is the mean field fit discussed in the text.

\item $\chi^{4/7}$ vs. coupling for the data plotted in Fig. 14. The dashed
line is the
prediction of the dimensional reduction scenerio.

\item Order parameter squared vs. coupling for the NJL model on $6 \times 36^2$
lattice
with four flavors. The dashed line is the fit discussed in the text.

\item Same as Fig. 16 except for the inverse susceptibility.

\item Same as Fig. 15 for the four flavor model on a $6 \times 36^2$ lattice.

\item Same as Fig. 16 except on a $12 \times 72^2$ lattice.

\item Same as Fig. 14 except for the four flavor model.

\item Same as Fig. 15 except for the four flavor model.

\item Same as Fig. 21 except the fit tries a critical couling of $.758$.

\item Log-log plot of the order parameter vs. mass at criticality. The dashed
line is
given by mean field theory, while the dotted line is given by the 2-d Ising
model.

\end{enumerate}

\end{document}